\documentclass{llncs}

\usepackage{graphicx}
\usepackage{wrapfig}

\newcommand{\outputs}{\mathsf{outputs}}
\newcommand{\sset}[2]{\left\{~#1  \left|
      \begin{array}{l}#2\end{array}
    \right.     \right\}}

\newcommand{\algorithm}[1]{\mbox{\sc\tt{#1}}}

\begin{document}

\title{Formalizing Size-Optimal Sorting Networks:\\Extracting a Certified Proof Checker}
\author{Lu\'{i}s Cruz-Filipe \and Peter Schneider-Kamp}
\authorrunning{L.~Cruz-Filipe and P.~Schneider-Kamp}
\institute{Dept. Mathematics and Computer Science,
  Univ. of Southern Denmark\\
  Campusvej 55, 5230 ODENSE M, Denmark\\
  \email{$\{$lcf,petersk$\}$@imada.sdu.dk}
}

\maketitle

\begin{abstract}
Since the proof of the four color theorem in 1976, computer-generated proofs have become a reality in mathematics and computer science. During the last decade, we have seen formal proofs using verified proof assistants being
used to verify the validity of such proofs.

In this paper, we describe a formalized theory of size-optimal sorting networks. From this formalization we extract a certified checker that successfully verifies computer-generated proofs of optimality on up to $8$~inputs. The checker relies on an untrusted oracle to shortcut the search for witnesses on more than 1.6 million NP-complete subproblems.
\end{abstract}

\section{Introduction}

Sorting networks are hardware-oriented algorithms to sort a fixed number of inputs using a predetermined sequence of comparisons between them.
They are built from a primitive operator -- the \emph{comparator} --, which reads the values on two channels, and interchanges them if necessary to guarantee that the smallest one is always on a predetermined channel.
Comparisons between independent pairs of values can be performed in parallel, and the two main optimization problems one wants to addressed are: how many comparators do we need to sort $n$ inputs (the \emph{optimal size} problem); and how many computation steps do we need to sort $n$ inputs (the \emph{optimal depth} problem).

So far, most results obtained on these two problems require exhaustive analysis of huge space states.
For example, in order to prove that the known sorting network of size~$25$ on $9$ inputs is optimal, one has to show that no sequence of $24$ comparators will sort \emph{all} sequences of $9$ inputs.
There are several theoretical results that help reduce the full search space of over $10^{37}$ possibilities, but the resulting number of possibilities is still daunting.

In previous work~\cite{ourICTAIpaper} we proposed a generate-and-prune algorithm to show size optimality of sorting networks, and used it to establish optimality of the $25$-comparator sorting networks on $9$ inputs.
This program includes a pruning step that takes two sequences of comparators and decides whether one of them can be ignored; however, this requires iteration over all $9!$ permutations of $9$ elements.
This expensive step, which often fails, has to be repeated billions of times throughout the whole run; therefore, the total execution required more than 10 years of computational time, or around one month on a state-of-the-art parallel cluster able to run $288$ threads simultaneously.

\pagebreak

During execution, we took care to record the comparator sequences and permutations that allowed \emph{successful} pruning steps.
Using this log, the whole run could be replicated and independently validated in just a few hours of sequential computation by shortcutting the expensive search process. While this additional step increased our confidence in the result, there remain reasonable grounds to argue whether these results constitute a valid mathematical proof.

In this paper we describe a Coq formalization of the relevant theory of sorting networks, from which we extract a certified checker with the goal of confirming the validity of our result. In order to obtain feasible runtimes, we bypass the expensive search process in the extracted certified checker by means of an oracle providing the information required for the pruning steps. The checker takes a skeptical approach towards the oracle: if the oracle provides wrong information at any step, the checker ignores it.
The oracle is implemented independently by reading the original log filento the data structures expected by the checker.

This paper focuses on the formalization of the theory underlying the checker.
We discuss how we can exploit the constructiveness of the theory to simplify the formalization and
identify the places where skepticism towards the oracle becomes relevant.
Finally, the interactive process of formalization itself revealed minor gaps in the hand-written proofs of \cite{ourICTAIpaper}, underlining the importance of formal proof assistants in computer-aided mathematical proofs.

This paper is structured as follows.
Section~\ref{sec:background} summarizes the theory of sorting networks relevant for this formalization, and describes the generate-and-prune algorithm together with the information logged during its execution.
Section~\ref{sec:defns} describes the formalization of the theory, with emphasis on the challenges encountered.
Section~\ref{sec:extraction} deals with the aspects of the formalization that have a direct impact on the extracted program, namely the specification of the oracle and the robustness needed to guarantee that unsound oracles do not compromise the final results.
Section~\ref{sec:execution} addresses the implementation of the oracle and the execution of the extracted code.
Finally, Section~\ref{sec:conclusions} presents some concluding remarks and directions in which this work can be extended.

\subsection{Related work}
The proof of the four colour theorem from 1976~\cite{KW1977,KWK1977} was not the first computer-assisted proof, but it was the first to generate broad awareness of a new area of mathematics, sometimes dubbed ``experimental'' or ``computational'' mathematics. Since then, numerous theorems in mathematics and computer science have been proved by computer-assisted and computer-generated proofs. Besides obvious philosophical debates about what constitutes a
mathematical proof, concerns about the validity of such proofs have been raised since. In particular, proofs based on exhausting the solution space have been met with skepticism.

During the last decade, we have seen an increasing use of verified proof assistants to create formally verified computer-generated proofs. This has been a success story, and it has resulted in a plethora of formalizations of mathematical proofs, a list too long to even start mentioning particular instances. \emph{Pars pro toto}, consider the formal proof of the four colour theorem from 2005 \cite{GG2008}.

Outside the world of formal proofs, computer-generated proofs are flourishing, too, and growing to tremendous sizes. The proof of Erd\"{o}s' discrepancy conjecture for $C= 2$ from 2014 \cite{DBLP:conf/sat/KonevL14} has been touted as one of the largest mathematical proofs and produced approx.~13~GB of proof witnesses.
Such large-scale proofs are extremely challenging for formal verification. Given the current state of theorem provers and computing equipment, it is unthinkable to use Claret et al.'s approach~\cite{DBLP:conf/itp/ClaretGRZ13} of importing an \emph{oracle} based on the proof witnesses into Coq, a process clearly prohibitive for such large-scale proofs as we consider.

The last years have seen the appearance of \emph{untrusted oracles}, e.g.\ for a verified compiler \cite{DBLP:journals/cacm/Leroy09} or for polyhedral analysis \cite{DBLP:conf/sas/FouilheMP13}. Here, the verified proof tool is relegated to a checker of the computations of the untrusted oracle, e.g., by using hand-written untrusted code to compute a result and verified (extracted) code to check it before continuing the computation.

The termination proof certification projects IsaFoR/CeTA \cite{DBLP:conf/itp/Thiemann13}, based on Isabelle/HOL, and A3PAT~\cite{DBLP:conf/rta/ContejeanCFPU11}, based on Coq, go one step further, and use an \emph{offline} untrusted oracle approach, where different termination analyzers provide proof witnesses, which are stored and later checked.
However, a typical termination proof has $10$-$100$ proof witnesses and totals a few KB to a few MB of data, and recent work~\cite{DBLP:journals/corr/SternagelT14} mentions problems were encountered when dealing with proofs using ``several hundred megabytes'' of oracle data.
In contrast, our target proof of size-optimality of sorting networks with $9$ inputs requires dealing with nearly $70$ million proof witnesses, totalling $27$ GB of oracle data.

\section{Background}
\label{sec:background}

A \emph{comparator network} $C$ with $n$ channels and size $k$ is a
sequence of \emph{comparators} $C = (i_1,j_1);\ldots;(i_k,j_k)$, where
each comparator $(i_\ell,j_\ell)$ is a pair of channels $1\leq i_\ell< j_\ell\leq n$.
If $C_1$ and $C_2$ are comparator networks with
$n$~channels, then
$C_1;C_2$ denotes the comparator network obtained by concatenating
$C_1$ and $C_2$.
An input $\vec x=x_1\ldots x_n\in\{0,1\}^n$ propagates through $C$ as
follows: $\vec x^0 = \vec x$, and for $0<\ell\leq k$, $\vec x^\ell$ is
the permutation of $\vec x^{\ell-1}$ obtained by interchanging $\vec
x^{\ell-1}_{i_\ell}$ and $\vec x^{\ell-1}_{j_\ell}$ whenever $\vec
x^{\ell-1}_{i_\ell}>\vec x^{\ell-1}_{j_\ell}$.  The output of the
network for input $\vec x$ is $C(\vec x)=\vec x^k$, and
$\outputs(C)=\sset{C(\vec x)}{\vec x\in\{0,1\}^n}$.  The comparator
network $C$ is a \emph{sorting network} if all elements of
$\outputs(C)$ are sorted (in ascending order).
\begin{wrapfigure}[6]{r}{0.25\textwidth}
\vspace*{-4ex}
  \quad$(a)$~\raisebox{-2ex}{\includegraphics{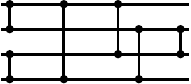}}

  \vspace{1em}

  \quad$(b)$~\raisebox{-2ex}{\includegraphics{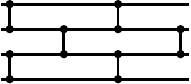}}
\end{wrapfigure}
The zero-one principle~\cite{Knuth73} implies that a sorting
network also sorts 
sequences over any other totally ordered set, e.g.~integers.
Images~$(a)$ and~$(b)$ on the right depict sorting networks on 4
channels, each consisting of 6 comparators.  The channels are
indicated as horizontal lines (with channel $4$ at the bottom),
comparators are indicated as vertical lines connecting a pair of
channels, and input values propagate from left to
right. The sequence of comparators associated with a picture
representation is obtained by a left-to-right, top-down
traversal. For example, the networks depicted above are: $(a)$
$(1,2); (3,4); (1,4); (1,3); (2,4); (2,3)$ 
and $(b)$~$(1,2); (3,4); (2,3); (1,2);(3,4); (2,3)$.

\subsection{Optimal Size Sorting Networks}%

The optimal-size sorting network problem is about finding the smallest
size, $S(n)$, of a sorting network on $n$ channels.  In~1964,
Floyd and Knuth presented sorting networks of optimal size for $n\leq 8$
and proved their optimality~\cite{Knuth66}.
Their proof required analyzing huge state spaces by means of a computer program, and the combinatorial explosion involved implied that there was no further progress on this problem in the next fifty years, until our proof
that $S(9) = 25$~\cite{ourICTAIpaper}.
The best currently known upper and lower bounds for $S(n)$ are given in the following table.
\begin{center}
\begin{tabular}{l|c|c|c|c|c|c|c|c|c|c|c|c|c|c|c|c}
$n$ & 1 & 2 & 3 & 4 & 5 & 6  & 7  & 8  & 9  & 10 & 11 & 12 & 13 & 14 & 15 & 16
\\ \hline
upper bound for $S(n)$ & 0 & 1 & 3 & 5 & 9 & 12 & 16 & 19 & 25 & 29 & 35 & 39 & 45 & 51 & 56 & 60\\
lower bound for $S(n)$\,&\,0\,&\,1\,&\,3\,&\,5\,&\,9\,&\,12\,&\,16\,&\,19\,&\,25\,&\,29\,&\,33\,&\,37\,&\,41\,&\,45\,&\,49\,&\,53\,
\end{tabular}
\end{center}
Given an $n$ channel comparator network
$C=(i_1,j_1);\ldots;(i_k,j_k)$ and a permutation $\pi$ on
$\{1,\ldots,n\}$, $\pi(C)$ is the sequence
$(\pi(i_1),\pi(j_1));\ldots;(\pi(i_k),\pi(j_k))$.  Formally, $\pi(C)$
is not a comparator network, but rather a \emph{generalized comparator network}: a comparator
network that may contain comparators $(i,j)$ with $i>j$,
which order their outputs in descending order instead of ascending.
A generalized sorting network can always be transformed
into a (standard) sorting network with the same size and depth by means of a procedure we will call \emph{standardization}: if $C=C_1;(i,j);C_2$ is a generalized sorting network with $C_1$ standard and $i>j$, then standardizing $C$ yields $C_1;(j,i);C'_2$, where $C'_2$ is obtained by (i)~interchanging $i$ and $j$ and (ii)~standardizing the result.
We write $C_1 \approx C_2$ ($C_1$ is equivalent to $C_2$) iff there is a
permutation $\pi$ such that $C_1$ is obtained by standardizing $\pi(C_2)$.
The two networks~$(a)$ and~$(b)$ above are equivalent via the permutation
$(1\,3)(2\,4)$.

\begin{lemma}
\label{lem:subsumption}
If $C$ is a generalized sorting network, then standardizing $C$ yields a sorting network.
\end{lemma}

The proof of this result, proposed as Exercise~5.3.4.16 in \cite{Knuth73}, is an induction proof that requires manipulating permutations and reasoning about the cardinality of $\outputs(C)$.

\subsection{The generate-and-prune approach}

Conceptually, one could inspect all $n$-channel comparator networks of size $k$
one-by-one to determine if any of them is a
sorting network. However, even for small $n$ such a naive approach is
combinatorically infeasible.
There are $n(n-1)/2$ comparators on $n$ channels,
and hence $(n(n-1)/2)^k$ networks with $k$ comparators.
For $n=9$, aiming
to prove that there does not exist a sorting network with $24$
comparators would mean inspecting approximately $2.25\times10^{37}$
comparator networks.
Moreover, checking whether a comparator network is a sorting network
is known to be a co-NP complete problem~\cite{Parberry91}.

In \cite{ourICTAIpaper} we propose an alternative approach, \emph{generate-and-prune}, which
is driven just as the naive approach, but takes advantage of the
abundance of symmetries in comparator networks, formalized via the notion of \emph{subsumption}.
Given two comparator networks on $n$ channels $C_a$ and $C_b$,
we say that $C_a$ subsumes $C_b$, and write $C_a\preceq C_b$,
if there exists a permutation $\pi$ such that $\pi(\outputs(C_a)) \subseteq \outputs(C_b)$.
If we need to make $\pi$ explicit, we will write $C_a\leq_{\pi}C_b$.

\begin{lemma}[\cite{ourICTAIpaper}]
  \label{lem:outputs}
  Let $C_a$ and $C_b$ be comparator networks on $n$ channels, both of
  the same size, and such that $C_a\preceq C_b$. Then, if there exists
  a sorting network $C_b;C$ of size~$k$, there also exists a sorting
  network $C_a;C'$ of size~$k$.
\end{lemma}

The proof of this result is constructive: under the hypotheses, the standardization of $C_a;\pi^{-1}(C)$ is a sorting network.
Again, this requires extensive manipulation of permutations and reasoning about the elements of $\outputs(C_a;\pi^{-1}(C))$.

Lemma~\ref{lem:outputs} implies that, when adding a next comparator
in the naive approach, we do not need to consider all possible
positions to place it. In particular, we can omit networks that are subsumed by others.

The \emph{generate-and-prune} algorithm
iteratively builds two sets $R^n_k$
and $N^n_k$ of $n$ channel networks of size $k$.
First, it initializes $R^n_0$ to consist of a single
element: the empty comparator network. Then, it repeatedly applies two
types of steps, \algorithm{Generate} and \algorithm{Prune}%
, as follows.

\begin{enumerate}
\item\algorithm{Generate}: Given $R^n_k$, construct $N^n_{k+1}$ by adding one comparator to each
  element of $R^n_k$ in all possible ways.
\item\algorithm{Prune}: Given $N^n_{k+1}$, construct
  $R^n_{k+1}$ such that every element of $N^n_{k+1}$ is subsumed by an element of $R^n_{k+1}$.
\end{enumerate}
The algorithm stops when a sorting network is found, which will make $\left|R^n_k\right|=1$.

To implement \verb+Prune+, we loop on $N^n_k$ and check whether the current network is subsumed by any of the previous ones; if this is the case, we ignore it.  Otherwise, we add it to $R^n_k$ and remove any networks already in this set that are subsumed by it.
This yields a double loop $N^n_k$ where at each iteration we need to find out whether a subsumption exists -- which, in the worst case, requires looping through all $n!$ permutations.
For $n=9$, the largest set $N^n_k$ is $N^9_{15}$, with over $18$~million elements, and there are potentially $300\times 10^{12}$ subsumptions to test.

These algorithms are straightforward to implement, test
and debug. Our implementation from \cite{ourICTAIpaper} written in Prolog and can be applied to
reconstruct all of the known values for $S(n)$ for $n\leq 6$ in under
an hour of computation on a single core and, after several optimizations and parallelization as described in~\cite{ourICTAIpaper}, was able to obtain the new value of $S(9)$.

Soundness of generate-and-prune follows from the observation that $N^n_k$ (and $R^n_k$) are \emph{complete} for the optimal size sorting
network problem on $n$ channels: if
there exists an optimal size sorting network on $n$ channels, then there exists one
of the form $C;C'$ for some $C\in N^n_k$ (or $C\in R^n_k$), for every $k$.

\subsection{Checking the Proof using Proof Witnesses}

Even though we proved all of the mathematical claims underlying the design of the generate-and-prune algorithm
and carefully checked the correctness of our Prolog implementation, it is reasonable to question the validity of the final result.
In \cite{ourICTAIpaper} we turned to Barendregt and
Wiedijk's \emph{de Bruijn criterion}~\cite{Barendregt2005}: every computer-generated
proof should be verifiable by an independent small program (a ``checker'').
We therefore augmented our code to produce a log file of successful subsumptions during execution, and implemented an independent Java verifier that was able to re-check the result without needing to replicate the expensive search steps
in just over $6$ hours of computational time.\footnote{The logs and the Java
verifier are available from: \url{http://imada.sdu.dk/\~{}petersk/sn/}}

However, we may again question the validity of the checker, and enter into an endless loop of validations.
In this paper, we propose a different goal: obtain a correct-by-design checker by extracting it from a formalized proof of the theory of sorting networks.
The reason for aiming at extracting a checker, rather than the full generate-and-prune algorithm, is that by using the log as an (untrusted) oracle we again gain a speedup of several orders of magnitude, as we completely avoid all the search steps.%

In total, we have logged proof witnesses for approx.~70 million
subsumptions, yielding a 27 GB log file.
Developing a formalization allowing the extraction of an efficient checker that uses an untrusted oracle of this magnitude is an exciting challenge that, to the best of our knowledge, has not been tackled before.
We proceed in two stages.
First, we formalize the theory of optimal-size sorting networks directly following~\cite{Knuth73}, including the new results from~\cite{Parberry91,ourICTAIpaper}.
Then we implement generate-and-prune with an oracle in Coq, prove its soundness, and extract a certified checker.

\section{Formalizing sorting networks}
\label{sec:defns}

Formalizing the theory of sorting networks presents some challenges relating to the very finite nature of the domain.
All the relevant notions are parameterized on the number $n$ of inputs, and thus the domain for most concepts is the finite set $\{0,1,\ldots,n-1\}$.%

Directly working with this set in Coq is very cumbersome due to the ensuing omnipresence of proof terms -- every number required as argument has to be accompanied by a proof that it is in the adequate range.
Furthermore, these proof terms are completely trivial, since the order relations on natural numbers are all decidable.
Therefore, we chose to define all relevant concepts in terms of natural numbers, and define additional properties specifying that particular instances fall in the appropriate range.
For example, a comparator is simply defined as a pair of natural numbers:
{\small\begin{verbatim}
Definition comparator : Set := (prod nat nat).
\end{verbatim}}
\noindent and we define with predicates stating that a particular comparator is a (standard) comparator on $n$ channels:
{\small\begin{verbatim}
Definition comp_channels (n:nat) (c:comparator) :=
  let (i,j) := c in (i<n) /\ (j<n) /\ (i<>j).

Definition comp_std (n:nat) (c:comparator) :=
  let (i,j) := c in (i<n) /\ (j<n) /\ (i<j).
\end{verbatim}}

Likewise, the type \verb+CN+ of comparator networks is defined as \verb+list comparator+, and there are predicates stating that a comparator network spans $n$ channels, and that it is standard (which simply state that all comparators in the list have the corresponding comparator property).

Comparator networks act on binary sequences, and we define these as a dependent type, similar to \verb+Vector+.
{\small\begin{verbatim}
Inductive bin_seq : nat -> Set := 
  | empty : bin_seq 0
  | zero : forall n:nat, bin_seq n -> bin_seq (S n)
  | one : forall n:nat, bin_seq n -> bin_seq (S n).
\end{verbatim}}
We then define two operations \verb+get+ and \verb+set+, such that \verb+(get n s)+ returns the element ($0$ or $1$) in position \verb+n+ of \verb+s+, and \verb+(set n s k)+ sets position \verb+n+ of \verb+s+ to $0$, if \verb+k+ is $0$, and $1$ otherwise.
Setting an index larger than the length of a sequence leaves the sequence unchanged, while attempting to get the value in an index out of range returns $2$.
Since the context for the usage of these functions always ensures that these situations do not occur, these options are immaterial for the formalization.

A sequence is sorted if its first element is $0$ and the remaining sequence is sorted, or if it consists entirely of $1$s.

{\small\begin{verbatim}
Fixpoint all_ones (n:nat) (x:bin_seq n) : Prop := match x with
  | empty => True
  | zero _ _ => False
  | one _ y => all_ones _ y
  end.

Fixpoint sorted (n:nat) (x:bin_seq n) : Prop := match x with
  | empty _ => True
  | zero _ y => sorted _ y
  | one _ y => all_ones _ y
  end.
\end{verbatim}}

Sequences propagate through comparator networks as expected; a sorting network is a comparator network that sorts all inputs.

{\small\begin{verbatim}
Fixpoint apply (c:comparator) n (s:bin_seq n) :=
  let (i,j):=c in let x:=(get s i) in let y:=(get s j) in
    match (le_lt_dec x y) with
    | left _ => s
    | right _ => set (set s j x) i y
    end.

Fixpoint full_apply (S:CN) n (s:bin_seq n) := match S with
  | nil => s
  | cons c S' => full_apply S' _ (apply c s)
  end.

Definition sorting_network (n:nat) (S:CN) := 
  (channels n S) /\ forall s:bin_seq n, sorted (full_apply S s).
\end{verbatim}}

We also define an alternative characterization of sorting networks in terms of their sets of outputs and prove its equivalence to this one, but for space constraints we will not present it here.

\emph{A posteriori}, this characterization is actually quite close to the corresponding mathematical definition: the comparator network $\{(0,2);(2,5)\}$ is a network on $6$ channels, but it is also a network on $9$ or~$11$~channels, or indeed on $n$ channels for any $n\geq 6$.
So this implementation option does not only make the formalization easier, but it is also a good model of the way we think about these objects.

\subsection{Proof methodology}

As discussed above, we completely separate between objects and proof terms.
This option is dictated by the constructive nature of the theory of sorting networks.
The key results underlying the pruning step of the generate-and-prune algorithm are all of the form
\begin{quote}
  If there is a sorting network $N$, then there is another sorting network $N'$ such that\ldots
\end{quote}
and the published proofs of these results~\cite{ourICTAIpaper,Knuth73,Parberry91} \emph{all} proceed by explicitly describing how to construct $N'$ from $N$.
We formalize these results as operators in order to simplify their reuse: instead of proving statements of the form
$\forall N.\varphi(N) \to \exists N'.\psi(N')$
we first define some transformation $T$ and prove that
$\forall N.\varphi(N) \to \psi(T(N))$
from which we can straightforwardly prove the original statement.

Keeping the proof terms separated from the definitions and formalizing existential proofs as operators has several advantages:
\begin{itemize}
\item it is much easier to define these operators and then prove that they satisfy the required properties than including proof terms in their definition;
\item the hypotheses underlying the properties themselves become much more explicit -- e.g.~renumbering channels yields a network on the same number of channels and of the same size, and these two results are independent;
\item later, we can use the operators directly together with the relevant lemmas, rather than repeatedly applying inversion on an existential statement;
\item additional properties of the operators that are needed later are easy to add as new lemmas, instead of requiring changes to the original theorem;
\item we automatically get proof irrelevance, as proof terms are universally quantified in lemmas.
\end{itemize}

As an example, recall the definition of standardization: given a comparator network $C$, pick the first comparator $(i,j)$ for which $i>j$, replace it with $(j,i)$ and exchange $i$ with $j$ in all subsequent comparators, then iterate the process until a fixpoint is reached.
Standardization is the key ingredient in Lemma~\ref{lem:outputs}, and it is formalized using well-founded recursion as follows.
{\small\begin{verbatim}
Function std (S:CN) {measure length S} : CN := match S with
  | nil => nil
  | cons c S' => let (x,y) := c in
         match (le_lt_dec x y) with
         | left _ => (x[<]y :: std S')
         | right Hxy => (y[<]x :: std (permute x y) S'))
end end.
\end{verbatim}}

We then prove that standardizing a comparator network on $n$ channels yields a standard comparator network on $n$ channels.
This is not completely trivial, because of the permutation in the recursive step: permuting channel labels $x$ and $y$ preserves the total number channels $n$ only if $x<n$ and $y<n$.
While these hypotheses trivially hold ($x$ and $y$ are channels in an $n$-channel network), requiring the corresponding proof terms in the definition of comparator would make the definition of \verb!std! unreadable.
In later results, we can use the \verb!std! operator when needed.
There are in total seven different properties that are needed in different places in the formalization,
and all but two are proven independently.
If we did not have \verb!std! as an operator, they would have to be formalized together as one gigantic result stating that for every network $C$ there exists another network $C'$ satisfying the seven necessary properties.

\subsection{Permutations}

Permutations are an essential part of the formalization, as they are used extensively in the proof of Lemma~\ref{lem:outputs}.

Representing permutations in Coq is a challenging problem, and there are several common alternatives.
The standard library includes an inductive type stating that two lists are permutations of each other; but manipulating it is cumbersome.
Furthermore, we are not interested in developing a theory of permutations, but rather in proving results about comparator networks whose channels are renamed by means of a permutation of the numbers $0,\ldots,n-1$, which we will hereafter refer to as ``a permutation of $[n]$''.
Therefore we want a definition that makes it easy and efficient to apply permutations to objects.

For this reason, we chose to represent permutations as finite functions.
A permutation $P$ is a list of pairs of natural numbers, with the intended meaning that $(i,j)\in P$ corresponds to $P$ mapping $i$ into $j$.
We assume that $P$ does not change $i$ if there is no pair $(i,j)\in P$ -- this makes it much simpler to represent transpositions, which are the only permutations we need to represent explicitly in the formalization.
In order for $P$ to be a valid permutation of $[n]$, several conditions have to hold:
\begin{enumerate}
\item all pairs $(i,j)\in P$ must satisfy $i<n$ and $j<n$;
\item no number may occur twice either as the first or as the second element of distinct pairs in $P$; and
\item the sets of numbers occuring as first or second elements of the pairs in $P$ must coincide.
\end{enumerate}

As before, we separate the datatype of permutations from the property of being a permutation.
Here, \verb+NoDup+ is the Coq standard library predicate stating that a list does not have duplicate elements, and \verb+all_lt(n,l)+ is an inductive predicate stating that all elements of \verb+l+ are smaller than \verb+n+.
{\small\begin{verbatim}
Definition permut := list (nat*nat).

Definition dom (P:permut) := map (fst (A:=nat) (B:=nat)) P.
Definition cod (P:permut) := map (snd (A:=nat) (B:=nat)) P.

Definition permutation n (P:permut) :=
  NoDup (dom P) /\ all_lt n (dom P) /\
  forall i, In i (dom P) <-> In i (cod P).
\end{verbatim}}
As a sanity-check, we prove the relationship with the permutations in the Coq standard library.
{\small\begin{verbatim}
Lemma permutation_Permutation : Permutation (dom P) (cod P).
\end{verbatim}}
All properties of permutations are added to the \verb+core+ hint database, so that Coq can automatically prove most properties of permutations required during the formalization.

We provide mechanisms to define permutations in four different ways, three of which correspond to the usage of permutations in proofs, and another one which will be necessary for interacting with the oracle.
The former are as follows.
\begin{enumerate}
\item The identity permutation is simply the empty list, and it is easily shown to be a permutation of $[n]$ for any $n$.
\item Given a permutation \verb+P+, we construct its inverse \verb+(inverse_perm P)+ by reversing all pairs in \verb+P+.  If \verb+P+ is a permutation of $[n]$, then so is \verb+(inverse_perm P)+.
\item The transposition \verb+(transposition i j)+ is the permutation that switches $i$ and $j$, leaving all other values unchanged.
  This transposition is defined as the list $\{(i,j),(j,i)\}$, and in order for it to be a permutation it is necessary that $i\neq j$ (otherwise the list $\{i,j\}$ contains duplicate elements).
  This condition therefore shows up in some results about transpositions; it is never a problem, though,
as all transpositions arise from the \verb+standardization+ function, where $i$ and $j$ are obtained from a comparator $(i,j)$.
\end{enumerate}

We also need to get permutations from the oracle, and here we use a different representation for efficiency reasons.
The log files record permutations as their output on the set $[n]$, so for example the transposition over $[4]$ exchanging $0$ and $2$ would be represented as $\{2,1,0,3\}$.
Since we want to be skeptic about the oracle, we do not assume anything about the lists we are given; rather, we show that the property of a list of natural numbers corresponding to a permutation on $[n]$ is decidable.
We then define a function \verb+make_perm+ to translate lists of natural numbers into (syntactic) permutations, and show that the resulting object satisfies \verb+permutation+ if the original list corresponds to a permutation.
{\small\begin{verbatim}
Variable n:nat.
Variable l:list nat.

Definition pre_perm := NoDup l /\ all_lt n l /\ length l = n.

Lemma pre_perm_dec : {pre_perm} + {~pre_perm}.

Lemma pre_perm_lemma : pre_perm -> permutation n (make_perm l).
\end{verbatim}}

Thus, our checker will be able to get a list of natural numbers from the oracle, test whether it corresponds to a permutation, and in the affirmative case use this information.

One might wonder whether we could not have represented permutations uniformly throughout.
The reason for not doing so is that we have two distinct objectives in mind.
While formalizing results, we are working with an unknown number $n$ of channels, and it is much simpler to represent permutations by only explicitly mentioning the values that are changed, as this allows for uniform representations of transpositions and the identity permutation.
Also, computing the inverse of a permutation is very simple with the finite function representation, but not from the compact list representation given by the oracle.
When running the extracted checker, however, we are concerned with efficiency.
The oracle will provide information on millions of subsumptions, so it is of the utmost importance to minimize its size.

\section{Formalizing generate-and-prune}
\label{sec:extraction}

Soundness of generate-and-prune algorithm relies on the notion of a complete set of filters.
When formalizing this concept, we needed to make two changes: the element $C$ of $R$ being extended must be a standard comparator network with no redundant comparators\footnote{Comparator $(i,j)$ in comparator network $C;(i,j);C'$ is redundant if $\vec x_i<\vec x_j$ for all $\vec x\in\outputs(C)$ -- in other words, $(i,j)$ never changes its inputs.};
and (as a consequence) the size of the sorting network extending $C$ is \emph{at most} $k$, since there is an upper bound on how many non-redundant comparisons we can make on $n$ inputs.%

{\small\begin{verbatim}
Definition size_complete (R:list CN) (n:nat) := forall k:nat,
 (exists C:CN, sorting_network n C /\ length C = k) ->
  exists C' C'':CN, In C' R /\ standard n (C'++C'')
        /\ (forall C1 c C2, (C'++C'') = (C1++c::C2) -> ~redundant n C1 c)
        /\ sorting_network n (C'++C'') /\ length (C'++C'') <= k.
\end{verbatim}}

These changes were discovered during the formalization of the original soundness proof~\cite{ourICTAIpaper}, which implictly used on the fact that the elements of the complete sets of filters constructed were not redundant.

We prove that the set $\{\emptyset\}$ is complete, and that if there is a complete set of filters $R$ whose elements all have size $k$, then all sorting networks on $n$ channels have size at least $k$.
This key property does not hold for the previous informal definition of size completeness.

{\small\begin{verbatim}
Lemma empty_complete : forall n, size_complete (nil::nil) n.

Lemma complete_size : forall R n k, size_complete R n ->
                     (forall C, In C R -> length C = k) ->
                      forall S, sorting_network n S -> length S >= k.
\end{verbatim}}

\subsection{The generation step}

The formalization of the generation step proceeds in two phases.
First, we define the simple function adding a comparator at the end of a comparator network in all possible ways, and \verb+Generate+ simply maps it into a set.
The function \verb+all_st_comps+ produces a list of all standard comparators on $n$ channels.

{\small\begin{verbatim}
Definition add_to_all (cc:list comparator) (C:CN) :=
  map (fun c => (C ++ (c :: nil))) cc.

Fixpoint Generate (R:list CN) (n:nat) := match R with
  | nil => nil
  | cons C R' => (add_to_all (all_st_comps n) C) ++ Generate R' n
  end.
\end{verbatim}}

Then, we use the fact that redundancy of the last comparator is decidable to define an optimized version that removes redundant networks.

{\small\begin{verbatim}
Definition last_red (n:nat) (C:CN) : Prop :=
  exists C' c, redundant n C' c /\ C = (C' ++ c :: nil).

Lemma last_red_dec : forall n C, {last_red n C} + {~last_red n C}.

Fixpoint filter_nred (n:nat) (R:list CN) := match R with
  | nil => nil
  | (C :: R') => match last_red_dec n C with
                 | left _ => filter_nred n R'
                 | right _ => C :: filter_nred n R'
end end.

Definition OGenerate (R:list CN) (n:nat) := filter_nred n (Generate R n).
\end{verbatim}}%

Both \verb+Generate+ and its optimized version map size complete sets into size complete sets, as long as the input set does not already contain a sorting network (in which case \verb+OGenerate+ would return an empty set).

{\small\begin{verbatim}
Theorem OGenerate_complete : forall R n, size_complete R n ->
                            (forall C, In C R -> ~sorting_network n C) ->
                                       size_complete (OGenerate R n) n.
\end{verbatim}}

The extracted code for these two functions coincides with their Coq definition, since there are no proof terms involved, and matches the pseudo-code in~\cite{ourICTAIpaper}.

\subsection{The pruning step}
\label{sec:pruning step}

For the pruning step, we need to work with the untrusted oracle.
We define an oracle to be a list of subsumption triples $\langle C,C',\pi\rangle$, with intended meaning that $C\leq_\pi C'$.
Using the oracle, we then define the pruning step as follows.

{\small\begin{verbatim}
Function Prune (O:Oracle) (R:list CN) (n:nat) {measure length R}
               : list CN := match O with
  | nil => R
  | cons (C,C',pi) O' => match (CN_eq_dec C C') with
      | left _ => R
      | right _ => match (In_dec CN_eq_dec C R) with
          | right _ => R
          | left _ => match (pre_perm_dec n pi) with
              | right _ => R
              | left A => match (subsumption_dec n C C' pi' Hpi) with
                  | right _ => R
                  | left _ => Prune O' (remove CN_eq_dec C' R) n
end end end end end.
\end{verbatim}}

The successive tests in \verb+Prune+ verify that: $C\neq C'$; $C\in R$; $\pi$ is a permutation; and $C\leq_\pi C'$.
If any of these fail, this subsumption is skipped, else $C'$ is removed from $R$.
For legibility, we wrote \verb+pi'+ for the actual permutation generated by \verb+pi+ and \verb+Hpi+ for the proof term stating that this is indeed a permutation.
The Java verifier from~\cite{ourICTAIpaper} did not validate that the permutations in the log files were correct permutations.

The key result states that this is a mapping from complete sets of filters into complete sets of filters, \emph{regardless of the correctness of the oracle}, as long as the input set contains only standard comparator networks with no redundant comparators, and all are of the same size.

{\small\begin{verbatim}
Theorem Prune_complete : forall O R n, size_complete R n ->
  (forall C, In C R -> standard n C) ->
  (forall C C' c C'', In C R -> C = C'++c::C'' -> ~redundant n C' c) ->
  (forall C C', In C R -> In C' R -> length C = length C') ->
  size_complete (Prune O R n) n.
\end{verbatim}}

This implementation is simpler than the pseudo-code in~\cite{ourICTAIpaper}, as the oracle allows us to bypass all search steps -- both for permutations and for possible subsumptions.

\subsection{Coupling everything together}

We now want to define the iterative generate-and-prune algorithm and prove its correctness.
Here we deviate somewhat from the original presentation.
Our algorithm will receive as inputs two natural numbers (the number of channels $m$ and the number of iterations $n$) and return one of three possible answers: \verb+(yes m k)+, meaning that a sorting network of size $k$ was found and that no sorting network of size $(k-1)$ exists; \verb+(no m k R H1 H2 H3)+, meaning that \verb+R+ is a set of standard (\verb+H3+) comparator networks of size $k$ (\verb+H2+), with no duplicates (\verb+H1+); or \verb+maybe+, meaning that an error occurred.
The proof terms in \verb+no+ are necessary for the correctness proof, but they are all removed in the extracted checker.
They make the code quite complex to read, so we present a simplified version where they are omitted.

{\small\begin{verbatim}
Inductive Answer : Set :=
  | yes : nat -> nat -> Answer
  | no : forall n k:nat, forall R:list CN, NoDup R ->
                        (forall C, In C R -> length C = k) ->
                        (forall C, In C R -> standard n C) -> Answer
  | maybe : Answer.

Fixpoint Generate_and_Prune (m n:nat) (O:list Oracle) := match n with
  | 0 => match m with
         | 0 => yes 0 0
         | 1 => yes 1 0
         | _ => no m 0 (nil :: nil) _ _ _
         end
  | S k => match O with
      | nil => maybe
      | X::O' => let GP := (Generate_and_Prune m k O') in
           match GP with
           | maybe => maybe
           | yes p q => yes p q
           | no p q R _ _ _ => let GP' := Prune X (OGenerate R p) p in
                               match (exists_SN_dec p GP' _) with
                                     | left _ => yes p (S q)
                                     | right _ => no p (S q) GP' _ _ _
end end end end.
\end{verbatim}}
Note the elimination over \verb+exists_SN_dec+, which states that we can decide whether a set contains a sorting network.
{\small\begin{verbatim}
Lemma exists_SN_dec : forall m l, (forall C, In C l -> channels m C) ->
                     {exists C, In C l /\ sorting_network m C} +
                     {forall C, In C l -> ~sorting_network m C}.
\end{verbatim}}

The correctness of the answer is shown in the two main theorems: if the answer is \verb+(yes m k)+, then the smallest sorting network on $m$ channels has size $k$; and if the answer is \verb+(no m k)+, then there is no sorting network on $m$ channels with size $k$ or smaller.
These results universally quantify over \verb+O+, meaning that they hold regardless of whether the oracle gives right or wrong information.
{\small\begin{verbatim}
Theorem GP_yes : forall m n O k, Generate_and_Prune m n O = yes m k ->
                (forall C, sorting_network m C -> length C >= k) /\
                 exists C, sorting_network m C /\ length C = k.

Theorem GP_no : forall m n O R HR0 HR1 HR2,
                     Generate_and_Prune m n O = no m n R HR0 HR1 HR2 ->
                     forall C, sorting_network m C -> length C > n.
\end{verbatim}}

The full Coq formalization and its generated documentation are available from \url{http://imada.sdu.dk/\~{}petersk/sn/}

\section{Running the extracted program}
\label{sec:execution}

We extracted the certified checker to Haskell using Coq's extraction mechanism. The result is a file
\verb+Checker.hs+ containing among others a Haskell function \verb+generate_and_Prune :: Nat -> Nat -> (List Oracle) -> Answer+.
In order to run this extracted certified checker, we wrote a Haskell interface
that calls \verb+generate_and_Prune+ function with the number of channels, the maximum size of the networks, and the list of the
oracle information, and then prints the answer.
This interface includes conversion functions from Haskell integers to the extracted naturals and a function implementing the oracle, as well as a definition of \verb+Checker.Answer+ to be an instance of the type class \verb+Show+ for printing the result.

It is important to stress that we do \emph{not} need to worry about soundness of almost any function defined in the interface, as the oracle is untrusted.
For example, a wrong conversion from natural numbers to their Peano representation will not impact the correctness of the execution (although it will definitely impact the execution \emph{time}, as all subsumptions will become invalid).
We only need to worry about the function printing the result, but this is straightforward to verify.

The extracted checker was able to validate the proofs of optimal size up to and including $n = 8$ in around one day -- roughly the same time it took to produce the original proof, albeit without search.
This required processing more than $300$~MB of proof witnesses for the roughly $1.6$~million subsumptions.
To the best of our knowledge, it constitutes the first formal proof of the results in~\cite{Knuth66}.
Experiments suggested that the verification of the proof that $S(9)=25$ would take around 20~years.%

\section{Conclusions}
\label{sec:conclusions}
We have have presented a formalization of the theory of size-optimal sorting networks, extracted a verified checker for size-optimality proofs, and used it to show that informal results obtained in previous work are correct.

Our main contribution is a \emph{formalization} of the theory of size-optimal sorting networks, including
an intuitive and reusable formalization of comparator networks and a new representation of permutations more suitable for computation.

Another immediate contribution is a \emph{certified checker}
that directly confirmed all the values of $S(n)$ quoted in~\cite{Knuth66}.

Subsequently, we focused on making the checker more efficient without changing the formalization described in this paper~\cite{ourextraction}.
This eventually allowed us to verify also the claim that $S(9)=25$.
We plan to apply the same technique -- first formalize, then optimize -- to other computer-generated proofs where formal verification has been prohibitively expensive so far.

\section*{Acknowledgements}
We would like to thank Femke van Raamsdonk, whose skepticism about our informal proof inspired this work, and Michael Codish for his support and his enthusiasm about sorting networks.
The authors were supported by the Danish Council for Independent Research, Natural Sciences. Computational resources were generously provided by the Danish Center for Scientific Computing.

\bibliographystyle{plain}
\bibliography{formalization}

\end{document}